\documentclass[12pt]{article}
\usepackage{mathptmx,amsfonts,amsmath,amssymb}
\usepackage{Iopconf}
\usepackage{graphicx}
\usepackage{bm}

\newcommand{\V}{\mathbf{V}}
\newcommand{\pr}{\text{pr}}

\begin{document}
\title{Discretizations preserving all Lie point symmetries of the
      Korteweg-de Vries equation}
\author{Francis Valiquette\dag \footnote{E-mail:
      valiquet@crm.umontreal.ca}}
\affil{\dag\ Centre de Recherches Math\'ematiques, Universit\'e de
      Montr\'eal, C.P. 6128, succ. Centre-ville, Montr\'eal, QC, H3C
      3J7, Canada} 

\beginabstract
We show how to descritize the Korteweg-de Vries (KdV)
equation in such a way as to preserve all the Lie point symmetries of the
continuous dif\mbox{}ferential equation.  It is shown that, for a
centered implicit scheme, there are at least two possible ways of doing so.
\endabstract

\section{Introduction}

Symmetries play an important role in our understanding of nature, they
are intrinsic and fundamental features of dif\mbox{}ferential equations in
mathematical physics.  Hence, they should be retained when discrete
analogs of such equations are constructed.  The aim of this paper is
to give a systematic approach for discretizing partial 
dif\mbox{}ferential equations (PDEs), involving one function 
of two variables $u(x,t)$, such that 
all the Lie point symmetries of the original problem are preserved.  
To illustrate the procedure we investigate the case of the 
KdV equation. We will show how it is possible to obtain invariant implicit
schemes.  Explicit schemes can also be derived but they often
exhibit numerical instability.  Finally, a numerical
application is realized using the schemes obtained.

\section{Discretization procedure}

The discretization of a PDE involving one depend variable $u$ and two
independant ones, $x$ and $t$, consists of sampling,
couples $(x,t)$ in the real plane.  Each point is 
labelled by a set of discrete indices:
\begin{equation}
(x_{m,n},t_{m,n}),\qquad m,n \in\mathbb{Z}.
\end{equation} 
The discretization of the independant variables induces a natural 
discretization of the dependant one
\begin{equation}
u_{m,n}=u(x_{m,n},t_{m,n}).
\end{equation}

A two-variable partial dif\mbox{}ference scheme (P$\Delta$S) approximating
the original PDE will be a set of three equations
relating the quantities $\{x,t,u\}$ at a finite number of points
\begin{equation}\label{PDS}
\left.\begin{aligned}
& E_k(\{x_{m+i,n+j},t_{m+i,n+j},u_{m+i,n+j}\})=0,\\
& 1\leq k\leq 3,\qquad i_1\leq i\leq i_2,\qquad j_1\leq j\leq j_2,
\end{aligned}\right.
\end{equation}
such that, in the continous limit, all three equations \eqref{PDS} reduce to
the original PDE.  By choice, we suppose that $E_{1,2}=0$
describe the lattice and impose that in the continous limit these
equations go to the identity $0=0$ while $E_3=0$ goes to the PDE.  
We also require that the system \eqref{PDS} be invariant under the same
group of symmetries as the original PDE.

\subsection{Lie point symmetries of P$\Delta$S}

Let $G$ be the group of Lie point symmetries, of order $N$,
for a given PDE
\begin{equation}
\left.\begin{aligned}
x^*_l&=X_l(x,t,u,\epsilon)=x+\epsilon \xi^l(x,t,u)+\Or (\epsilon^2) \\
t^*_l&=T_l(x,t,u,\epsilon)=t+\epsilon \tau^l(x,t,u)+\Or (\epsilon^2),
\qquad l=1,\ldots,N \\
u^*_l&=U_l(x,t,u,\epsilon)=u+\epsilon
\phi^l(x,t,u)+\Or (\epsilon^2).
\end{aligned}\right.
\end{equation}
To this group is associated the Lie algebra of vector fields
\begin{equation}\label{algebra}
\V_l=\xi^l(x,t,u)\partial_x+\tau^l(x,t,u)\partial_t+
     \phi^l(x,t,u)\partial_u,\qquad l=1,\ldots,N.
\end{equation}

As in the continous case, we define a prolongation of \eqref{algebra}, but
in a dif\mbox{}ferent fashion.  The prolongation is realized by requiring that
the vector field acts at all points figuring in \eqref{PDS}:
\begin{equation}
\pr\V_l:=\sum_{i=m+i_1}^{m+i_2}\sum_{j=n+j_1}^{n+j_2}[\xi_{ij}^l
\partial_{x_{ij}}+\tau_{ij}^l\partial_{t_{ij}}
+\phi_{ij}^l\partial_{u_{ij}}]
\end{equation}
where
\begin{displaymath}
\xi_{ij}^l=\xi^l(x_{ij},t_{ij},u_{ij}),\quad
\tau_{ij}^l=\tau^l(x_{ij},t_{ij},u_{ij})\quad
\text{and}\quad
\phi_{ij}^l=\phi^l(x_{ij},t_{ij},u_{ij}).
\end{displaymath}

Let $M$ be the manifold on which $G$ acts, 
i.e. $M\thicksim\{x_{m+i_1,n+j_1},t_{m+i_1,n+j_1},u_{m+i_1,n+j_1},
\ldots,$ $x_{m+i_2,n+j_2},t_{m+i_2,n+j_2},u_{m+i_2,n+j_2}\}$.  
The quantity $I:M\to \mathbb{R}$ is said to be
strongly invariant if it satisfies
\begin{equation}
\pr\V_l I =0,\qquad l=1,\ldots,N.
\end{equation}
Using the method of the characteristics, we obtain a set of 
elementary invariants
$\{I_1,\ldots,I_\alpha\}$. The
number of them is given by
\begin{equation}
\alpha=\text{dim }M-\text{rank } Z,\qquad \alpha\geq 0,
\end{equation}
where $Z$ is the $N\times (i_2-i_1+j_2-j_1)$ matrix
\begin{equation}
Z=\begin{pmatrix}
\xi^1_{m+i_1,n+j_1} & \tau^1_{m+i_1,n+j_1} & \phi^1_{m+i_1,n+j_1}
& \ldots & \xi^1_{m+i_2,n+j_2} & \tau^1_{m-+i_2,n+j_2} & 
\phi^1_{m+i_2,n+j_2}\\
\vdots & \vdots & \vdots  & \ddots & \vdots & \vdots & \vdots \\
\xi^N_{m+i_1,n+j_1} & \tau^N_{m+i_1,n+j_1} & \phi^N_{m+i_1,n+j_1}
& \ldots & \xi^N_{m+i_2,n+j_2} & \tau^N_{m+i_2,n+j_2} &
\phi^N_{m+i_2,n+j_2}
\end{pmatrix}.
\end{equation}

A difference equation will be strongly invariant under $G$ if it can 
be written as
\begin{equation}\label{strongly invariant}
E(I_1,\ldots,I_\alpha)=0.
\end{equation}
Other invariant difference equations can be obtain if 
there exist expressions 
\begin{displaymath}
E(x_{m+i_1,n+j_1},t_{m+i_1,n+j_1},u_{m+i_1,n+j_1},\ldots,
x_{m+i_2,n+j_2},t_{m+i_2,n+j_2},u_{m+i_2,n+j_2})=0
\end{displaymath}
such that the rank of $Z$ is not maximal.  Such equations are said to
be weakly invariant and satisfy
\begin{equation}
\pr\V_l E \bigg\lvert_{E=0}=0,\qquad l=1,\ldots,N.
\end{equation}

\section{Discretization of the Korteweg-de Vries equation}

It is well known \cite{Olver}, that the KdV equation
\begin{equation}
u_t=uu_x+u_{xxx}
\end{equation}
admits a 4-parameter Lie symmetry group of point transformations,
generated by the infinitesimal operators:
\begin{equation}\label{KdV algebra}
\begin{split}
\V_1=\partial_x,\qquad \V_2=\partial_t,\qquad
\V_3=t\partial_x-\partial_u, \\
\V_4=x\partial_x+3t\partial_t-2u\partial_u.\qquad\quad
\end{split}
\end{equation}

Before computing the strong invariants of \eqref{KdV algebra}, we point
out the fact that
\begin{equation}
T_+ \equiv t_{m,n+1}-t_{m,n}=0
\end{equation}
is a weakly invariant equation.  Hence, the strong invariants can be computed
on a scheme with flat time layers, which is physically desirable.  
For numerical stability reasons, we compute the invariants 
on an implicit scheme involving the points shown below

\setlength{\unitlength}{1mm} \begin{center}
\begin{picture}(130,50)
\qbezier(10,40)(55,40)(120,40)
\qbezier(55,10)(60,30)(70,40)
\put(55,10){\circle*{2}} 
\put(10,40){\circle*{2}} 
\put(40,40){\circle*{2}}
\put(70,40){\circle*{2}}
\put(90,40){\circle*{2}}
\put(120,40){\circle*{2}}
\put(-1,42){($\hat{x}_{--},\hat{t},\hat{u}_{--}$)}
\put(32,42){($\hat{x}_-,\hat{t},\hat{u}_-$)}
\put(65,42){($\hat{x},\hat{t},\hat{u}$)}
\put(82,42){($\hat{x}_+,\hat{t},\hat{u}_+$)}
\put(110,42){($\hat{x}_{++},\hat{t},\hat{u}_{++}$)}
\put(41,9){($x,t,u$)}
\put(69,23.5){$\tau$}
\put(70,27.5){\vector(0,1){11.5}}
\put(70,21.5){\vector(0,-1){11.5}}
\put(60.5,6){$\sigma$}
\put(59.5,7){\vector(-1,0){5}}
\put(64.5,7){\vector(1,0){5.5}}
\put(23,35){$\hat{h}_{--}$} 
\put(31,36){\vector(1,0){10}}
\put(21.5,36){\vector(-1,0){11.5}}
\put(53,35){$\hat{h}_-$}
\put(58,36){\vector(1,0){12}}
\put(51.5,36){\vector(-1,0){12}}
\put(80,35){$\hat{h}_+$}
\put(85,36){\vector(1,0){6}}
\put(78.5,36){\vector(-1,0){8.25}}
\put(104,35){$\hat{h}_{++}$}
\put(111,36){\vector(1,0){9}}
\put(102.5,36){\vector(-1,0){12}}
\put(110,10){\vector(1,0){10}}
\put(110,10){\vector(0,1){10}}
\put(121,9){$n$}
\put(109,21){$m$}
\end{picture}\end{center}
For notational simplicity, we write
\begin{displaymath}
(x_{m,n},t_{m,n},u_{m,n})\equiv(x,t,u),\;
(x_{m+1,n},t_{m+1,n},u_{m+1,n})\equiv(\hat{x},\hat{t},\hat{u}),\;
(x_{m,n\pm 1},t_{m,n\pm 1},u_{m,n\pm 1})\equiv(x_\pm
,t_\pm,u_\pm)
\end{displaymath}
and so on.  The results of the computations are: 
\begin{equation}
\begin{aligned}
&I_1=\frac{\hat{h}_+}{\hat{h}_-}\qquad\;\;
I_2=\frac{\hat{h}_{++}}{\hat{h}_+}\qquad\qquad\;\;
I_3=\frac{\hat{h}_-}{\hat{h}_{--}}\qquad\;\;
I_4=\frac{\hat{h}_+^3}{\tau_+}\\
&I_5=\frac{\sigma+\tau u}{\hat{h}_+}\qquad\qquad\qquad\qquad\quad
I_6=\hat{h}_+^2(\hat{u}_+-u)\\
&I_{7}=\tau \hat{u}_x^{--}=\tau 
\left( \frac{\hat{u}_--\hat{u}_{--}}{\hat{h}_{--}}\right)\qquad
I_{8}=\tau \hat{u}_x^-=\tau \left( \frac{\hat{u}-
\hat{u}_-}{\hat{h}_-}\right)\\
&I_{9}=\tau \hat{u}_x^+=\tau \left( \frac{\hat{u}_+ 
-\hat{u}}{\hat{h}_+}\right)\qquad\quad\;\;
I_{10}=\tau \hat{u}_x^{++}=\tau 
\left( \frac{\hat{u}_{++} -\hat{u}_+}{\hat{h}_{++}}\right)
\end{aligned}\label{invariants}
\end{equation}

From the set of invariants \eqref{invariants}, different type of
schemes can be obtained. We will restrict our attention to
schemes for which the evolution of the solution is given implicitly
over a lattice that can be determined explicitly. 
  
\subsection{Lattice with uniform steps in $x$}
 
\begin{gather}
T_+=0\nonumber\\
I_1=1\label{uniform inv}\\
I_{6}=\frac{1}{2}{(I_6+I_5I_4)(I_9+I_8)+
(I_{10}-I_9)-(I_8-I_7)}.\nonumber
\end{gather}
In terms of the original variables, system \eqref{uniform inv}
reduces to
\begin{gather}
T_+=0\nonumber\\
\hat{h}_+=\hat{h}_-\equiv\hat{h} \label{uniform inv exp}\\
\frac{(\hat{u}-u)}{\tau}=\hat{u}\frac{(\hat{u}_+-\hat{u}_-)}{2\hat{h}}+
\frac{(\hat{u}_{++}-2\hat{u}_++2\hat{u}_--\hat{u}_{--})}{2\hat{h}^3}+
\frac{\sigma}{\tau}
\frac{(\hat{u}_+-\hat{u}_-)}{2\hat{h}}.\nonumber
\end{gather}

The third equation of \eqref{uniform inv exp} converges in the
continuous limit to the KdV equation if
\begin{equation}
\left\lvert\frac{\sigma}{\tau}\right\rvert
\xrightarrow[\tau\to 0]{\sigma\to 0}
\lvert R \rvert < \infty.
\end{equation}
Hence $\sigma = \Or(\tau)$.
Further more, the equations for the lattice can be solved and give
\begin{equation}
\begin{aligned}
t_{m,n}&=\gamma(m),\\
x_{m,n}&=\alpha(m)n+\beta(m),
\end{aligned}
\end{equation}
where $\gamma(m)$, $\alpha(m)$ and $\beta(m)$ are arbitrary functions
that we can choose.  The extra term
\begin{displaymath}
\frac{\sigma}{\tau}\frac{(\hat{u}_+-\hat{u}_-)}{2\hat{h}}
\end{displaymath}
in the discrete KdV equation is geometrically seen to 
be a correction term to the solution taking in account the 
possible displacement of the points in $x$ as the time evolves.

\subsection{Lattice depending on the solution}

Schemes for which the evolution of the lattice depends on the solution
have already been obtained by Dorodnitsyn \etal, \cite{Dorodnitsyn}. 
They use the concept of Lagrangian coordinates which is a point of view
that differs from ours.  Nevertheless, we get

\begin{gather}
T_+=0\nonumber\\
I_5=0\\
I_6=\frac{6}{I_2+2+2I_1^{-1}+(I_3I_1)^{-1}}\left\{
\frac{I_{10}-I_9}{I_2+1}-\frac{I_8-I_7}{I_3^{-1}+1}I_1\right\}\nonumber
\end{gather}
which in terms of the original variables gives
\begin{gather}
T_+=0\nonumber\\
\sigma=-\tau u \label{inv exp}\\
\frac{\hat{u}-u}{\tau}=\frac{6}{\hat{h}_{++}+2\hat{h}_+
+2\hat{h}_-+\hat{h}_{--}}\left\{ \frac{\hat{u}_x^{++} -
\hat{u}_x^+}{\hat{h}_{++}+\hat{h}_+} - \frac{\hat{u}_x^- -
\hat{u}_x^{--}}{\hat{h}_{--}+\hat{h}_-}\right\}.\nonumber
\end{gather}

From the numerical point of view, the scheme \eqref{inv exp} is
interesting since there is no nonlinear term in the equation
governing the evolution of the solution.  Hence, the computation
time is reduced since we avoid nonlinear numerical methods.
The absence of the nonlinear term is compensated by the specific 
evolution of the lattice.

\section{Numerical application}

This section is not meant to be an exhaustive investigation of all
the numerical aspects of our invariant schemes.  Actually, the aim is 
to give an illustration of the usefulness of such schemes 
for computing numerical solutions.
  
The solution we have choosen to investigate is 
\begin{equation}\label{numerical solution}
u(x,t)=-\frac{x}{t}. 
\end{equation}
Now, if we want to use the scheme \eqref{uniform inv exp}, we must first 
specify the lattice over which the numerical solution is to be
computed.  From all the possible choices, we will look at two specific 
cases.  As a first choice, since \eqref{numerical solution} is linear
in $x$, we may take
\begin{equation}\label{evolutive lattice}
x_{m+1,n}-x_{m,n}\equiv\sigma=\tau\frac{x}{t}.
\end{equation}
The solution of \eqref{evolutive lattice} is given by
\begin{equation}
x_{m,n}=h_0\left( \frac{t+\tau}{t}\right)^m n
\end{equation}
where $h_0$ is the initial step in $x$.  With this choice, 
scheme \eqref{uniform inv exp} reduces to 
the scheme \eqref{inv exp}.  As a second choice, we choose $\sigma$
to be zero.  In that case, the lattice is orthogonal and the 
discretization reduces to the standard implicit discretization 
of the KdV equation.

\begin{figure}[t]
\begin{minipage}[t]{0.4\textwidth}
\centering
\includegraphics[height=4cm,width=7cm]{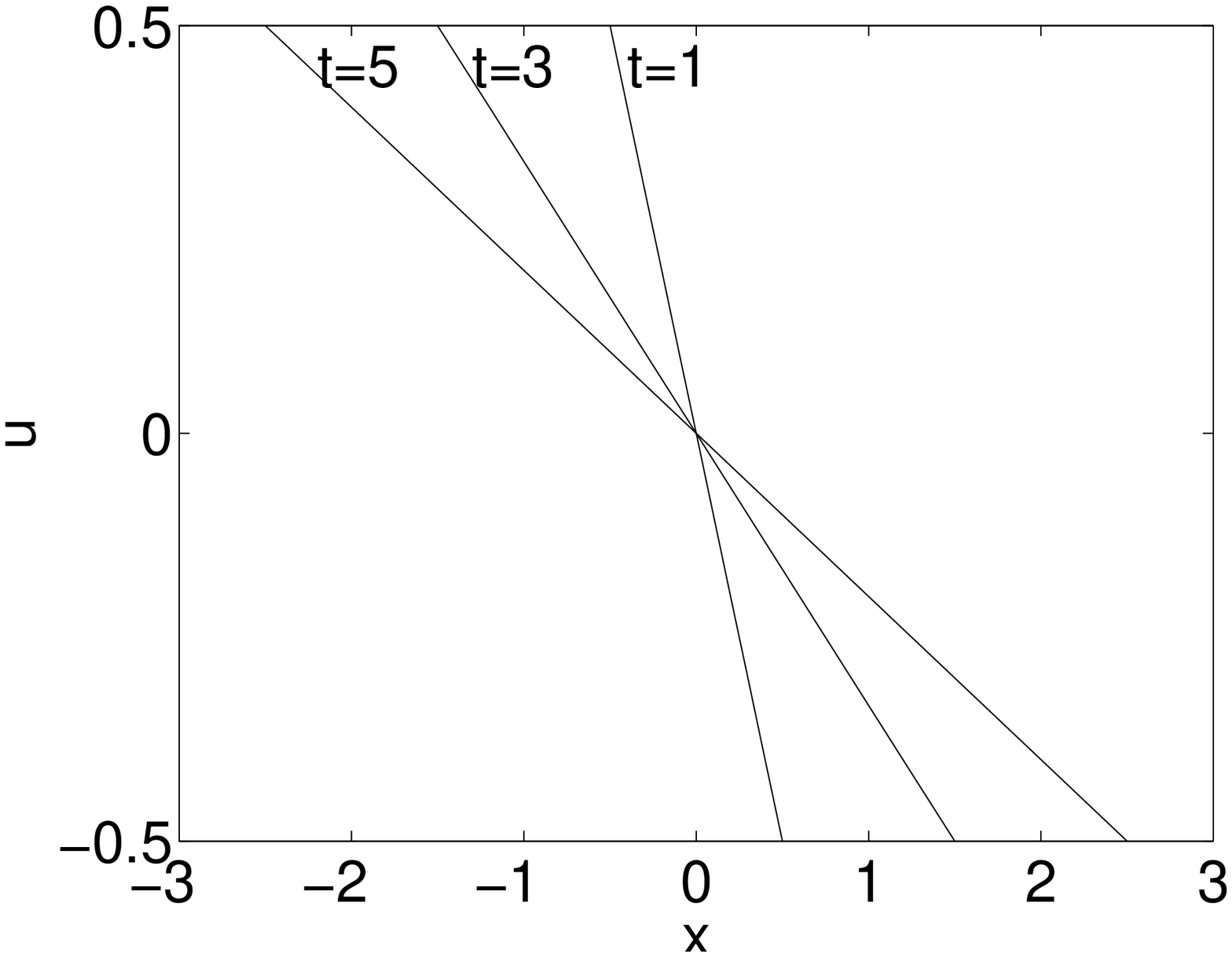}
\caption{Numerical solution on the evolutive lattice scheme.}
\label{solution}
\end{minipage}\hfill
\begin{minipage}[t]{0.4\textwidth}
\centering
\includegraphics[height=4cm,width=7cm]{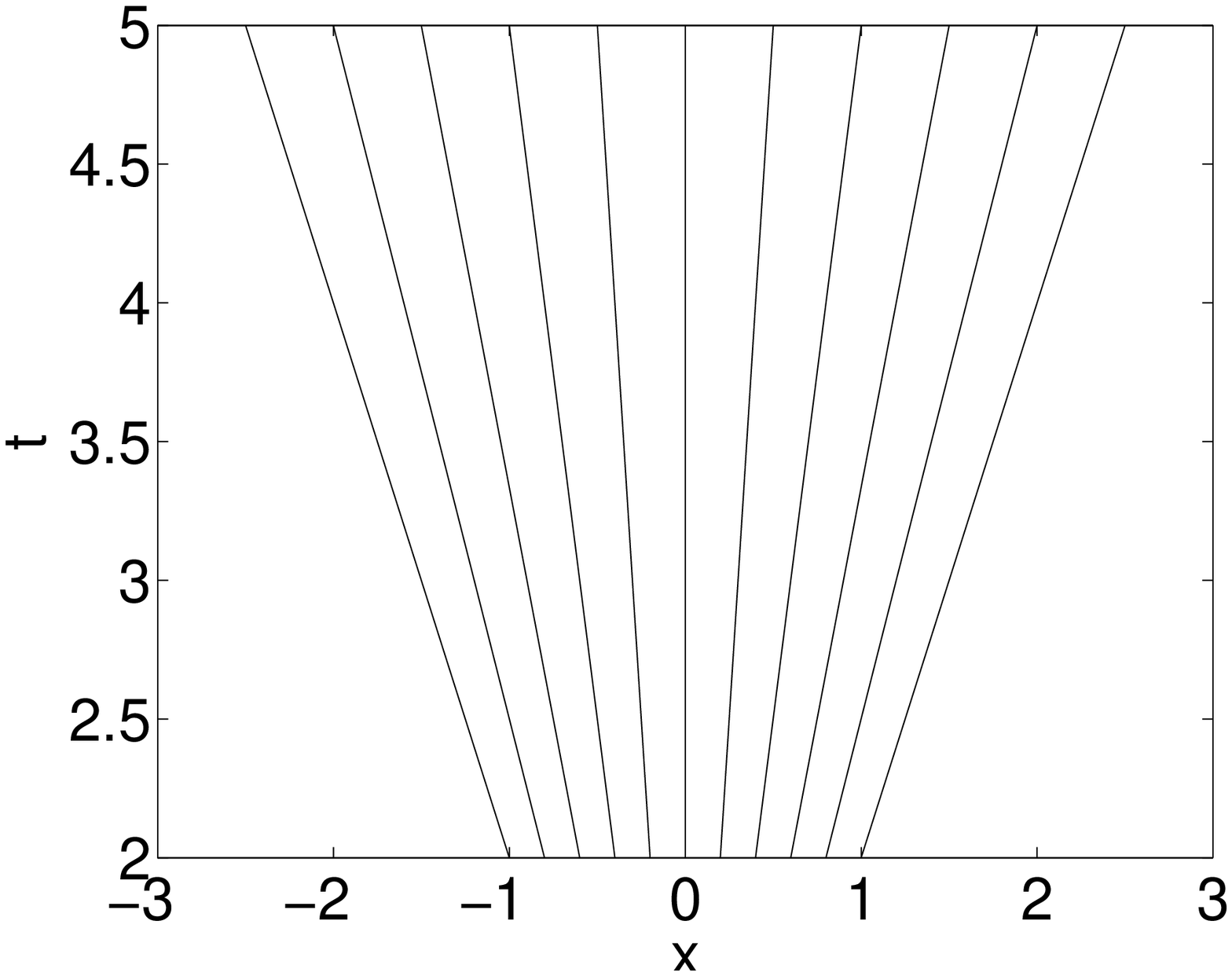}
\caption{Evolution of the lattice.}\label{lattice}
\end{minipage}
\end{figure}

\begin{figure}[h]
\begin{minipage}[t]{0.4\textwidth}
\centering
\includegraphics[height=4cm,width=7cm]{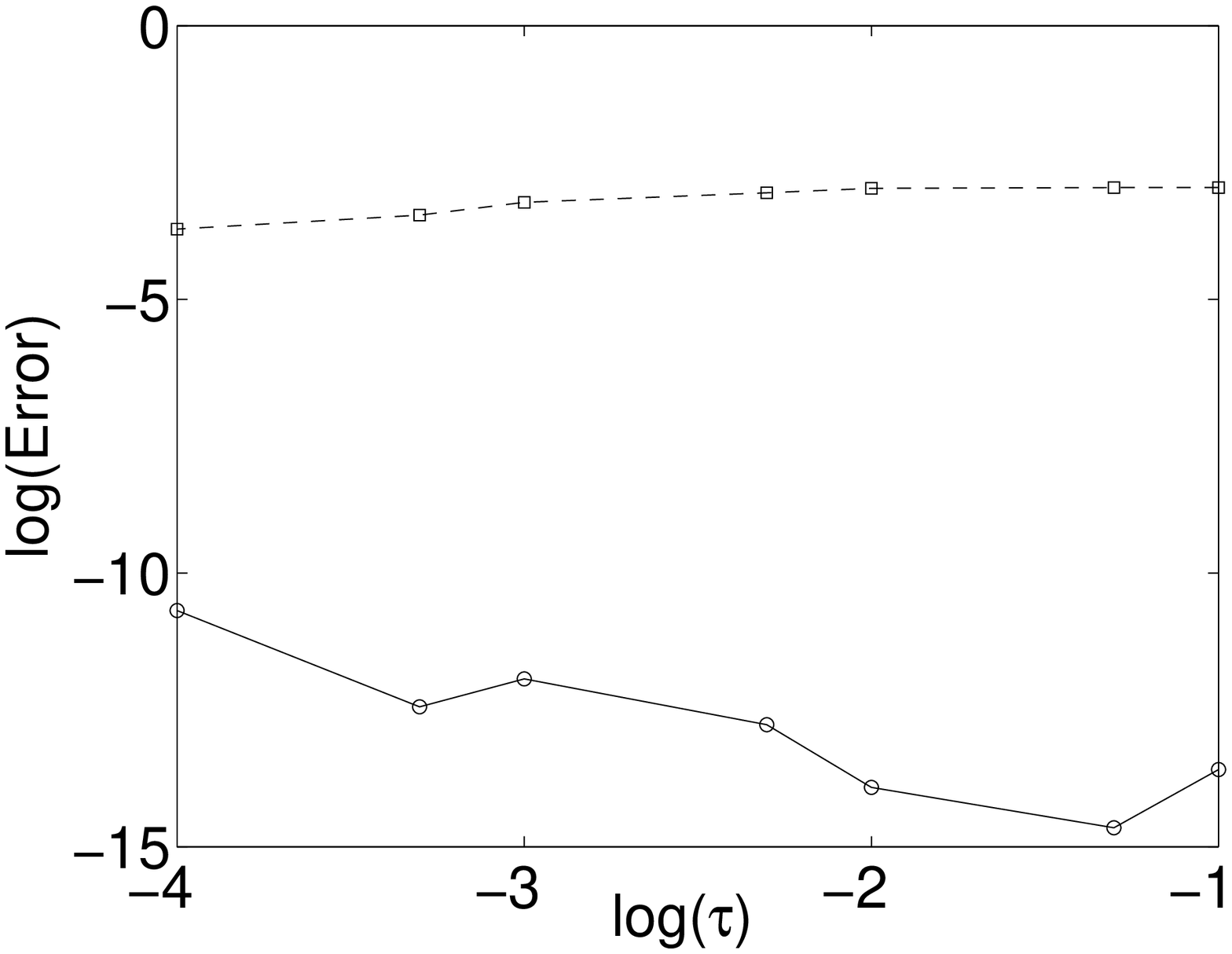}
\caption{Evolution of the error as a fonction of $\tau$:
$h(t_0)=0.1$, \full \;evolutive lattice, 
\broken \;orthogonal lattice.}\label{error_T}
\end{minipage}\hfill
\begin{minipage}[t]{0.4\textwidth}
\centering
\includegraphics[height=4cm,width=7cm]{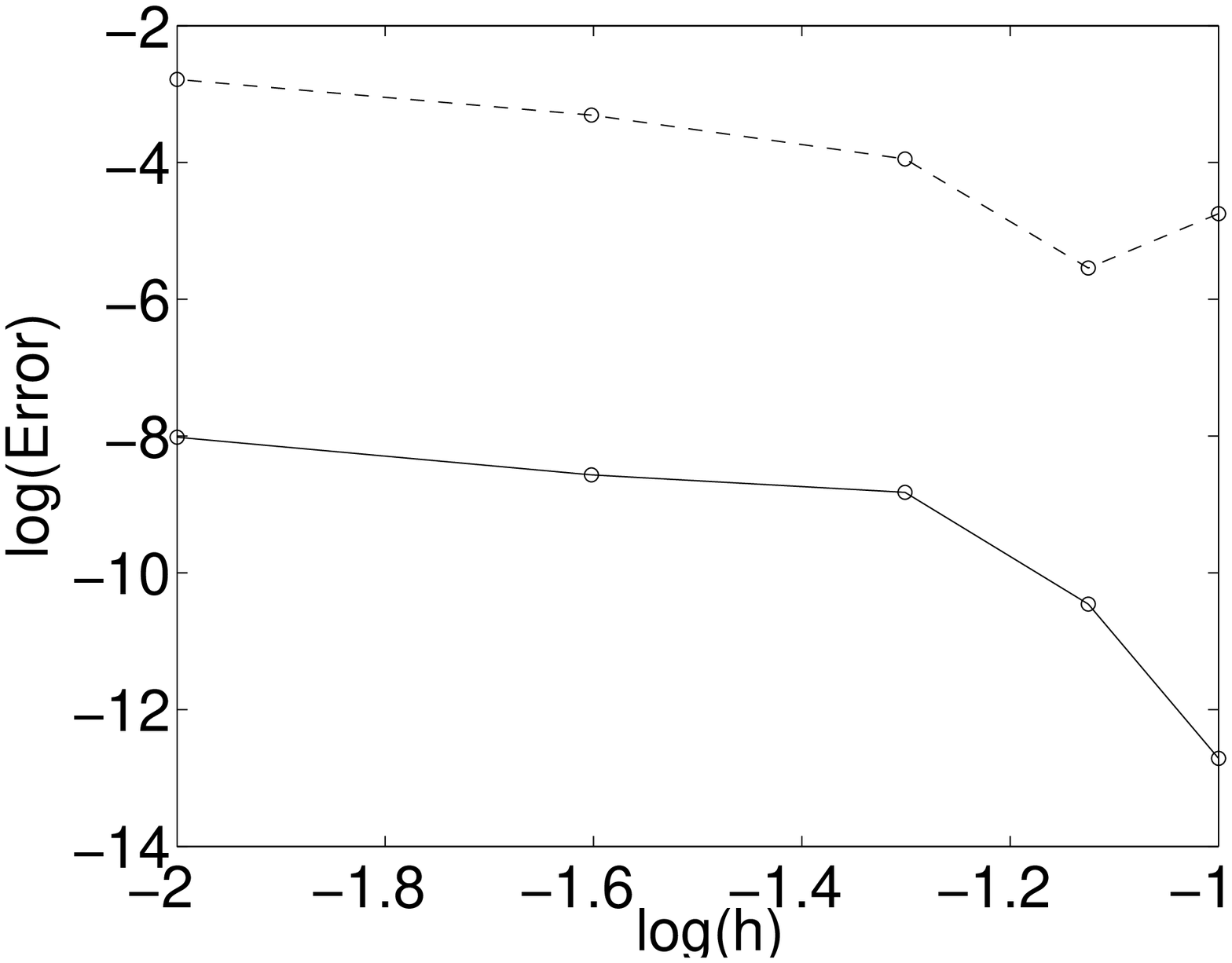}
\caption{Evolution of the error as a fonction of $h_0$:
$\tau=0.000001$ \full \;evolutive lattice, 
\broken \;orthogonal lattice.}\label{error_h}
\end{minipage}
\end{figure}

In the Figures \ref{solution} and \ref{lattice} we have plotted 
the evolution of the solution and lattice for the scheme define on the
evolutive grid. The initial step in $x$ was $h_0=0.1$ 
and in the time variable, we've taken a constant step, $\tau=0.1$.  

We've also plotted the evolution of the error as a fonction of the
step sizes in the figures \ref{error_T} and \ref{error_h} for the 
two schemes choosen.  The error has been computed using the $l_\infty$
norm.  As it can be seen, the error due to the evolutive lattice is
much smaller than the one generated with the orthogonal mesh. 

From this numerical simulation, we clearly see, that for the scheme
\eqref{uniform inv exp}, the choice of the lattice influences a lot
the numerical precision.  At the moment, we don't have a systematic
way of determining the lattice that will give the best numerical
results.  On the other hand, with the scheme \eqref{inv exp}
everything is determined by the system.  The only unknowns are the 
initial steps in $x$ and those in $t$.  
 
\section{Conclusion}

We have shown that it is possible to discretize the KdV on an
implicit scheme such as to preserve all the symmetries of the original
equation in at least two ways.  From the theoretical
point of view, these schemes are interesting since they preserve 
all the group properties of the original equation.  Hence, symmetry
reduction can hopefully be used to obtain exact solutions.  
From the numerical point of view, we see from our exemple,  
that the invariant schemes have the potential of giving better 
numerical results.

\section*{Acknowledgements}

I would like to thank Pavel Winternitz for all his 
judicious advice throught out the process of this work 
and NSERC for their financial support. 


\end{document}